\documentclass[letterpaper,12pt]{article}   
\usepackage{osajnl2} 

\def\aa{A\&A}
\def\aass{A\&A Sup. Ser.}

\begin{document}

\title{An analytical approach to optimize AC biasing of bolometers}


\author{Andrea Catalano,$^{1,2,*}$ Alain Coulais,$^2$ Jean-Michel
  Lamarre$^2$} \address{$^1$Laboratoire AstroParticule et Cosmologie
  (APC), Universit\'e Paris Diderot, CNRS/IN2P3, Observatoire de Paris, 10, rue Alice Domon et
  L\'eonie Duquet, 75205, Paris cedex 13, France}
\address{$^2$ LERMA, Observatoire de Paris et CNRS, 61 Avenue de
  l'Observatoire, 75014 Paris, France} \address{$^*$Corresponding
  author: catalano@apc.univ-paris7.fr}

\begin{abstract}

Bolometers are most often biased by Alternative Current (AC) in order to get rid of
low frequency noises that plague Direct Current (DC) amplification systems.  When stray
capacitance is present, the responsivity of the bolometer differs
significantly from the expectations of the classical theories. We
develop an analytical model which facilitates the optimization of the
AC readout electronics design and tuning. This model is applied to
cases not far from the bolometers in the Planck space mission.  We
study how the responsivity and the NEP (Noise Equivalent Power) of an AC biased bolometer depend on the essential parameters: bias current, heat sink temperature and
background power, modulation frequency of the bias, and stray capacitance. We show that
the optimal AC bias current in the bolometer is significantly
different from that of the DC case as soon as a stray capacitance is
present due to the difference in the electro-thermal feedback. We also
compare the performance of square and sine bias currents and show a
slight theoretical advantage for the last one.  This work resulted
from the need to be able to predict the real behaviour of AC biased
bolometers in an extended range of working parameters. It proved to be
applicable to optimize the tuning of the Planck High Frequency Instrument (HFI) bolometers.

\end{abstract}


\maketitle 

\section{Introduction}

Bolometers are now the most sensitive receivers for astrophysical
observations in the submillimetre spectral range. After decades of
improvement, they are able to operate with a sensitivity limited
by the photon noise of the observed source when operated outside of
the atmosphere \cite{Bock2009}. The principle of a bolometer is
that the heat deposited by the incoming radiation is measured by a
thermometer. The theory of bolometers has been developed in founding
papers \cite{Jones1953} and refined later
\cite{Mather1982,Mather1984a,Mather1984b}. They have shown that their
photometric responsivity strongly depends on its interaction with the
readout electronics, through the variation of the electrical power
deposited in the thermometer (the electro-thermal feedback). These
theories have been developed for a semiconductor thermometer element biased by a Direct Current (DC)
voltage through a load resistor. The readout electronics for bolometers experienced a
radical change more than a decade ago. Most of them are now using a
modulated bias current in order to get rid of low frequency noises
that plague amplification systems
\cite{Rieke1989,Wilbanks1990,Delvin1993,Gaertner1997,Kreysa2003}.

The theory developed for a DC bias must be altered for Alternative Current (AC) biased
bolometers in the presence of stray capacitance in the circuit. This
was evidenced in the Planck-HFI instrument \cite{Lamarre2010} in
spite of the fact that the readout electronics had been designed
\cite{Gaertner1997} to mimic, as far as possible, the operation of a
DC bias. Very significant differences were found in absolute responsivity
and even in value of the optimal bias current for the Planck bolometers. This
was shown to be mostly due to the effect of parasitic capacitances in
the wiring, which cannot be neglected in many practical experimental
setups. This effects have been studied in several papers dedicated to
the characterization of bolometers and calorimeters by measuring their
effective impedance (e.g. \cite{Vaillancourt2005}), but we are here
essentially interested in effective tools able to predict the responsivity
and optimise the tuning of bolometers in specific
configurations. Brute force modelling based on numerical integration
of thermal and electrical equations of the bolometers proved to be
feasible but computationally too heavy to be applied on wide ranges of
the many parameters of the models. To facilitate the computation, we
have developed an analytical model of the responsivity of AC biased
semiconductor bolometers. This model was used as an aid to predict
the behaviour of the bolometer of Planck-HFI and to optimize their
tuning. Its numerical application proved to be flexible and fast
enough to study the effects of all variable parameters.

This paper describes this analytical model and its application with a
set of parameters not too far from the realistic cases encountered in
Planck-HFI. The next section is dedicated to the differential
equations driving the thermal and the electrical behaviour of the
electro-thermal system comprising the bolometer and its readout
electronics. It focuses on the derivation of an analytical solution
giving the responsivity for both the DC and AC biased cases. Section three
addresses the various noises encountered and shows that the optimal
bias currents are different in the two cases. In the fourth section
the model is applied to analyze the effects of some essential
parameters (cold stage temperature, modulation frequency, value
of the stray capacitance, optical background). Section five deals with
the shape of the periodic bias wave to cover the case of square bias
current used in Planck-HFI.

\section{The Theoretical Model}

\subsection{Bolometer Model}

Let us consider a low temperature bolometer consisting of an absorber
attached to a semiconductor thermometer. The bolometer is attached to
a heat sink at temperature $T_0$ through a thermal link of thermal
conductance $G_s$. The incoming optical power deposits energy in the
absorber and heats the whole bolometer including the thermometer. The
absorbed optical power will determine the equilibrium temperature
$T_b$ of the bolometer:

\begin{equation}\label{bolo}
  G_s(T_b-T_0)=W_{tot}
\end{equation}

Where  $W_{tot}$ is the total power dissipated in the bolometer that is $W=P+Q$
where $Q$ is the absorbed radiant power and $P(t)=V(t)I(t)$ is the
electrical power.

$G_s$  can be well represented in many cases by:

$$G_s=G_{s0}(T_b/T_0)^{\beta}$$

where $G_{s0}$ is the static thermal conductance at temperature $T_0$
($100 mK$ in the case we investigate here). 

The dominant electrical conduction mechanism in the thermometer is the variable
range hopping between localised sites and the resistance of the device
varies with both applied voltage and temperature.

The relation between the resistance and the temperature of the
bolometer \cite{Piat2006} is set by :

\begin{equation}
  R(T,E)=R_* exp\left(\left(\frac{T_g}{T}\right)^n-\frac{eEL}{K_bT}\right)
\end{equation}

where $T_g$ is a characteristic parameter of the material, $R_*$ is a
parameter depending on the material and the geometry of the element,
$L$ is related to the average hopping distance and $E$ is the electric
field across the device.  In absence of electrical non-linearities and
other effects such as electron-phonon decoupling, the thermistor
resistance depends only on temperature:
\begin{equation}\label{resi}
  R(T)=R_* exp\left(\frac{T_g}{T}\right)^n
\end{equation}

The impedance changes induced by the temperature variations can
be measured by an appropriate readout circuit that we are going to
detail and discuss hereafter.

\subsection{Readout Electronics}

\begin{figure}[t!]
  \centering
  \includegraphics[width=8cm,keepaspectratio]{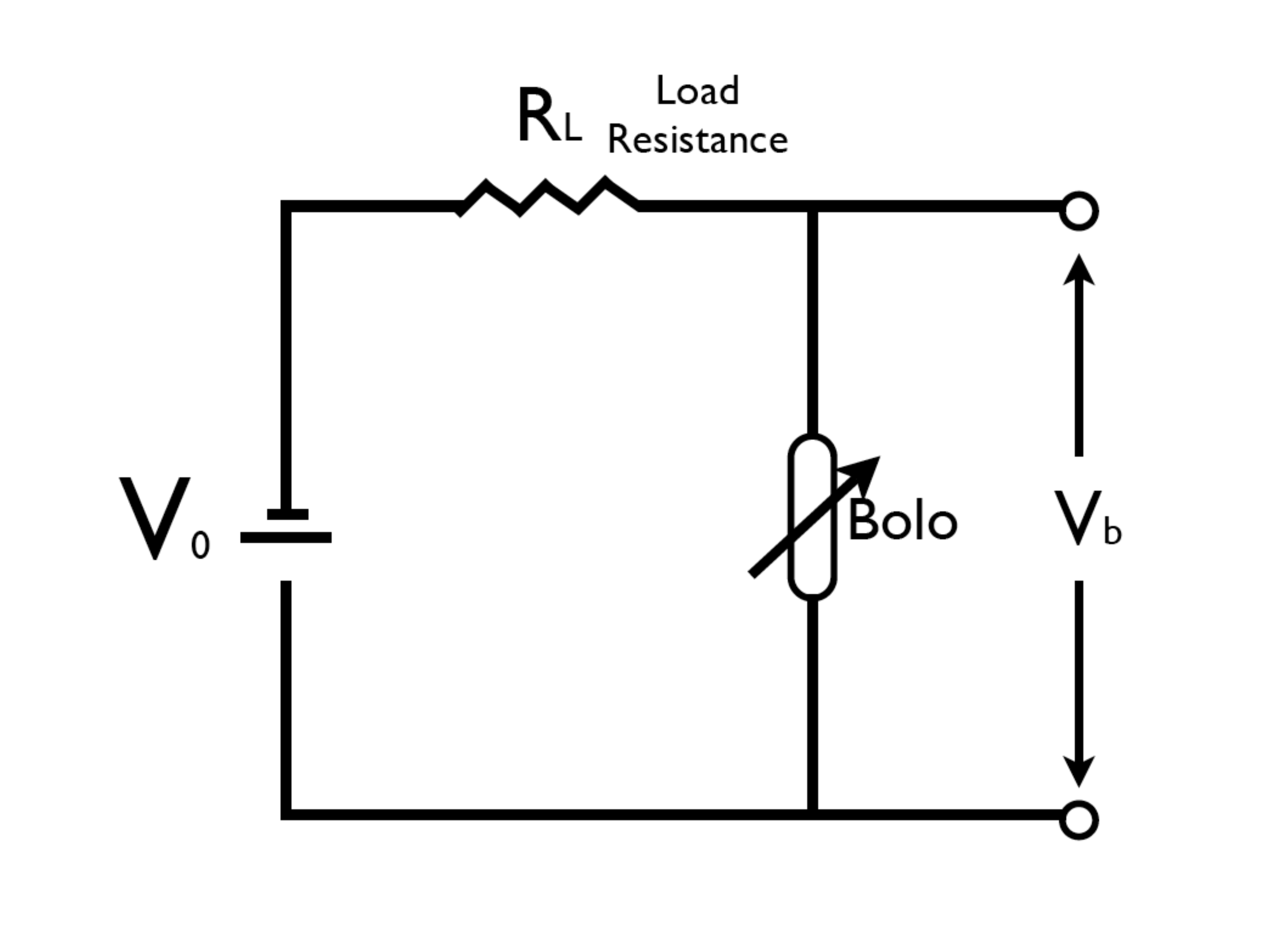}
  \includegraphics[width=8cm,keepaspectratio]{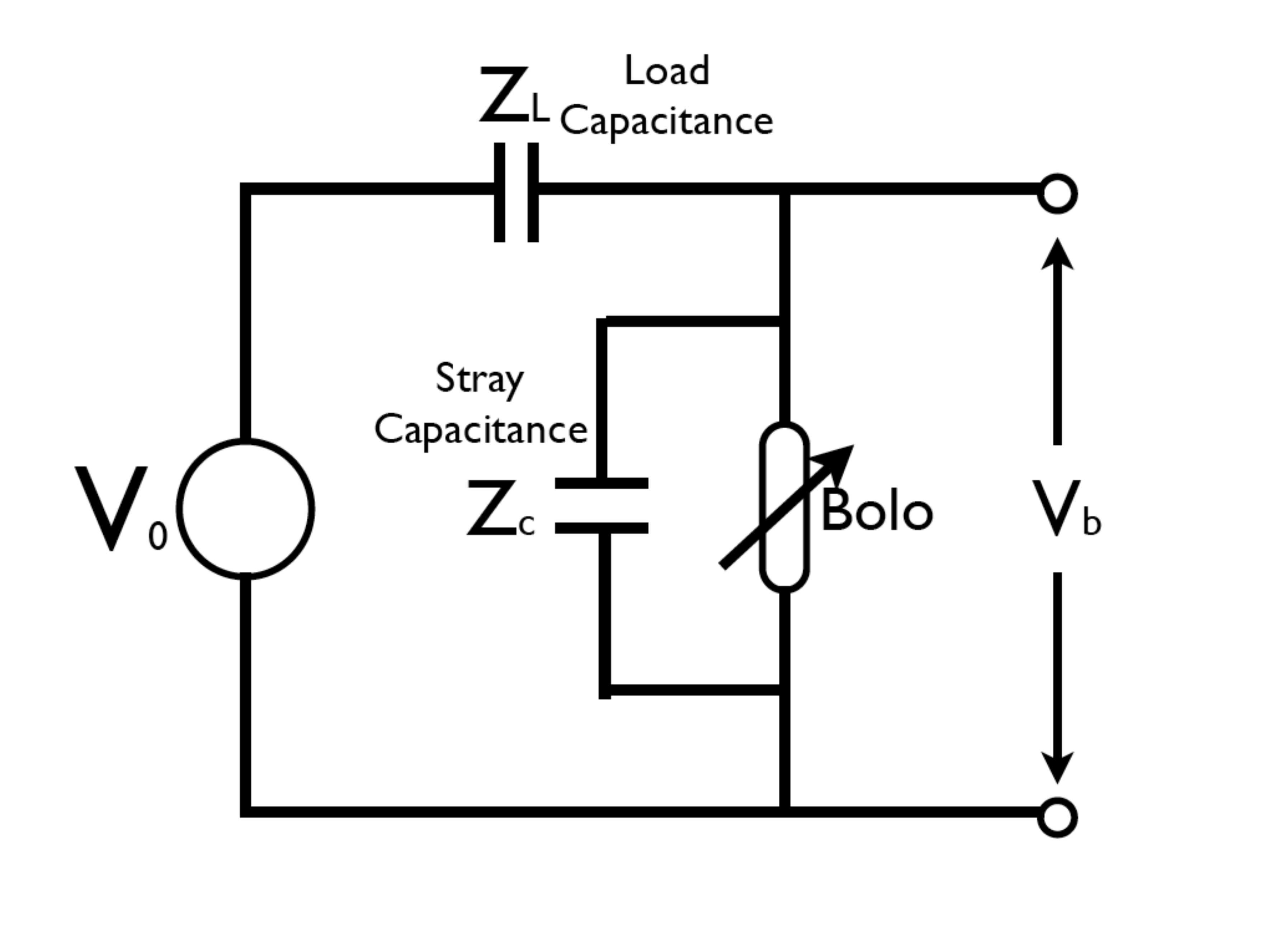}
  \caption{Schemes of a DC (left) and an AC (right) bias circuits}
  \label{fig:ACmod}
\end{figure}

The Readout Electronics is designed to measure the
impedance of the temperature sensitive element of the bolometer. This
is done by injecting a current, and therefore depositing power in
the bolometer, which changes its temperature. Consequently, the
bolometer responsivity and performance strongly depend on the design of
the readout electronics.

The Responsivity $ \Re$ is the derivative of the bolometer voltage with respect to the optical absorbed power $W$:

\begin{equation}
  \Re=\frac{dV}{dW}
\end{equation}

It is a strong indicator of the coupling efficiency achieved by readout electronics for a given bolometer.

We compare here after the responsivity obtained with a classical DC bias and a sine-shape AC bias.

\textbf{DC Responsivity : } the bolometer is biased with a DC bias
voltage through a load resistance $R_L$, and the voltage $V_b$ is
measured with an amplifier with an high input resistance (Fig
~\ref{fig:ACmod} left).  The general equation of a DC biased circuit
is :

\begin{equation}\label{equreu}
  V_b^{DC}=\frac{R_b}{R_b+R_L}V_0
\end{equation}

where $R_L$ and $R_b$ are the impedances of the load and the
bolometer. $V_0$ is the total input voltage.

The electrical responsivity $\Re_{el}$ can be written using the
Zwerling formalism \cite{Zwerdling1968}

\begin{equation}
  \Re_{el}=\frac{\alpha \varphi_{DC} R_{b} I_{b}}{G_{e}}
\end{equation}

where :

$$\varphi_{DC}=\frac{R_L}{R_b+R_L}$$

and $G_{e}$ is the equivalent thermal conductance :

\begin{equation}
  G_{e}= G_{s0}-\alpha R_{b} I_{b}^{2}(2 \varphi_{DC} -1)
\end{equation}

$\alpha$ is the temperature coefficient of resistance of the bolometer
:

\begin{equation}\label{eq2}
  \frac{dR_b}{dT_b}=\alpha \cdot R_b
\end{equation}

The advantage of this DC setup is the use of a well established
theory \cite{Jones1953,Mather1982}. The
optical power $W_{opt}$ absorbed by the bolometer and the responsivity of
the bolometer can be directly computed in the time domain.  On the other
hand, a DC bias current increases the level of low
frequency noise, like Flicker noise, making the detection of a
faint and slowly varying optical signal impossible. In addition, the Johnson noise
produced by the load resistor forces us to put this element on the
coldest cryogenic stage.

\textbf{AC Responsivity: } Let us consider now an AC bias circuit as
presented in Fig.~\ref{fig:ACmod} right.

The voltage at the ends of bolometer is:

\begin{equation}
  V_b^{AC}=V_0 \cdot \frac{R_b Z_c}{Z_L Z_c+R_b(Z_L+Z_c)}\label{aceq}
\end{equation}

If we assume that the AC bias frequency $F_{mod}$ is much higher than the
bolometer cut-off frequency ($F_{mod}>>\frac{G_e}{2 \pi C}$), then we
can consider only the average electrical power and a steady state
responsivity and neglect short term variations.  We can derive a modified
Zwerdling's formula for the responsivity by following the method used
in the previous section for a DC biased bolometer:

\begin{equation}
  \Re_{el}=\frac{\alpha \varphi_{AC} R_b I_b}{G^{AC}_e}
\end{equation}

where $G_{AC}$ the dynamic thermal conductance is equal to:
\begin{equation}
  G^{AC}_e=G_s+\frac{dG_s}{dT}(T-T_0)-\alpha R_b I_b^2 (2
  \varphi_{AC}-1)
\end{equation}

Here the $ \varphi_{AC}$ factor is:
\begin{equation}
  \varphi_{AC}=\frac{Z_L Z_c}{Z_L Z_c + R_b(Z_L + Z_c)}
\end{equation}

where, if $Z_L$ is a resistor:

$$ \displaystyle\lim_{Z_c\to\infty} \varphi_{AC} = \varphi_{DC}
$$

\begin{figure}[t!]
  \centering
  \includegraphics[width=8cm,keepaspectratio]{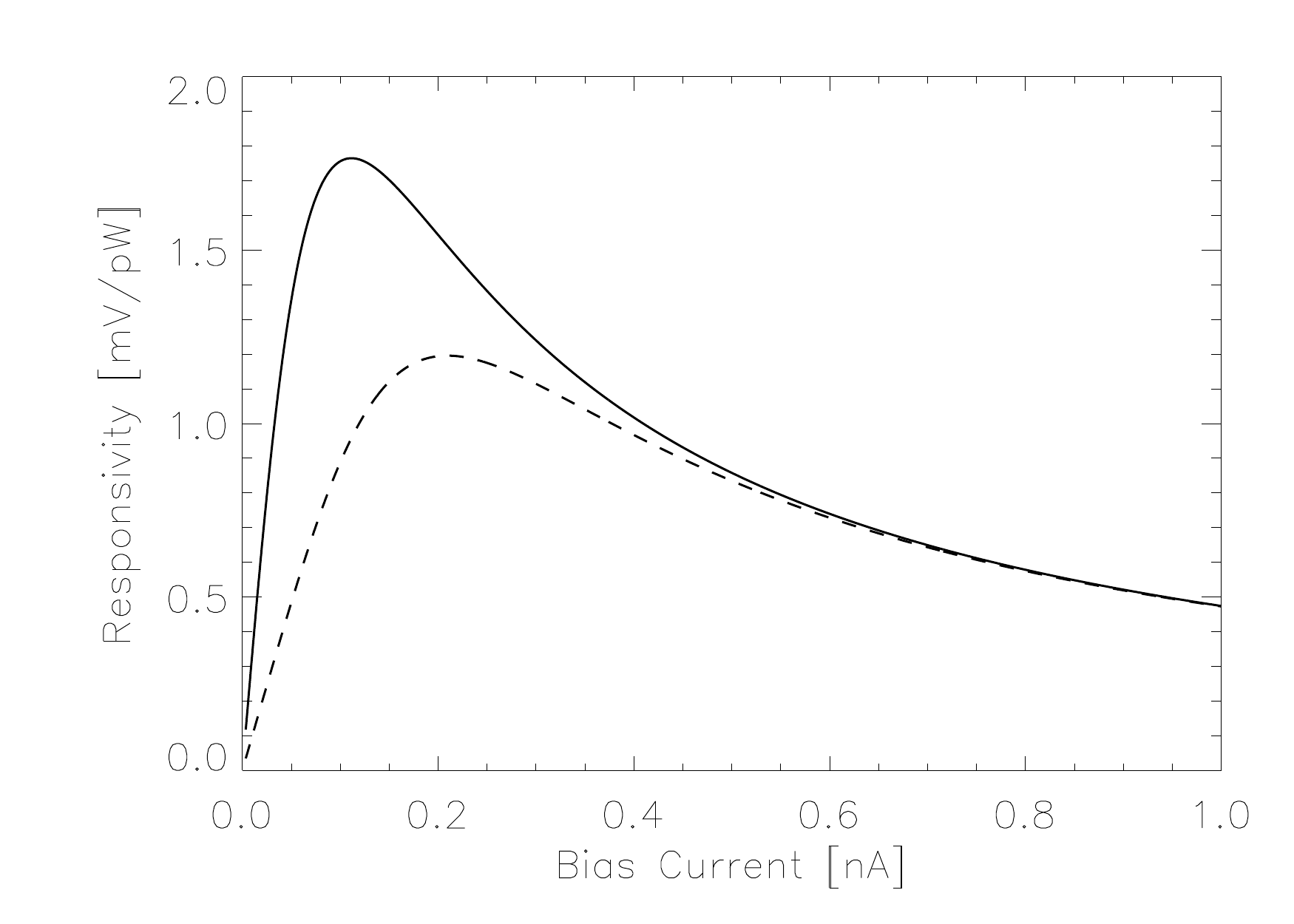}
  \includegraphics[width=8cm,keepaspectratio]{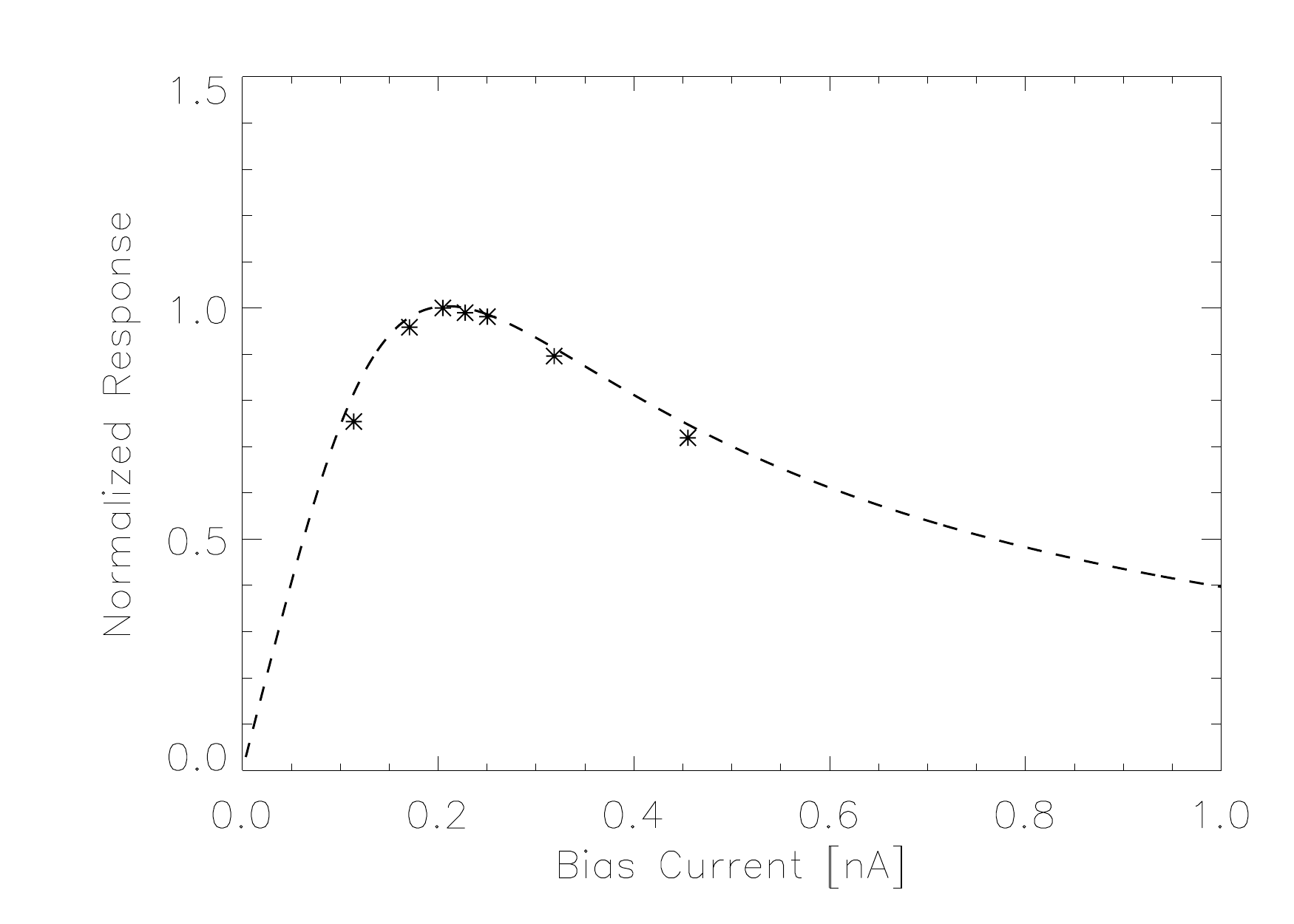}
  \caption{Left: Simulation of responsivity versus bias current in the bolometer for DC (solid curve) and AC
    sine (dashed curve) bias currents in the case of a 3 mm bolometer (using parameters from Tab. \ref{tab:tab}). For
    AC model we consider a sine wave bias with a stray capacitance
    of $C_p$ = 130 pF and plot the responsivity versus the r.m.s. value
    of the bias current. Right: Consistency of the AC model (dashed curve) with experimental measurements (stars points) taken from ground calibration of Planck HFI.  The disagreement is of the order of 1 \%.}
  \label{fig:comp}
\end{figure}

If we consider the module of the $\varphi$ factor we can plot
(Fig.~\ref{fig:comp}) a responsivity versus bias current for DC and AC
currents (sine-shape) bolometer. We can conclude that in terms of
responsivity a DC electronics is preferable. 
For an AC bias the maximum in responsivity
is lower and is obtained with higher bias current in the
bolometer. 
We show in the next section that for slowly varying signals, AC bias has a decisive advantage in sensitivity.

\begin{figure}[t!]
  \centering
  \includegraphics[width=8cm,keepaspectratio]{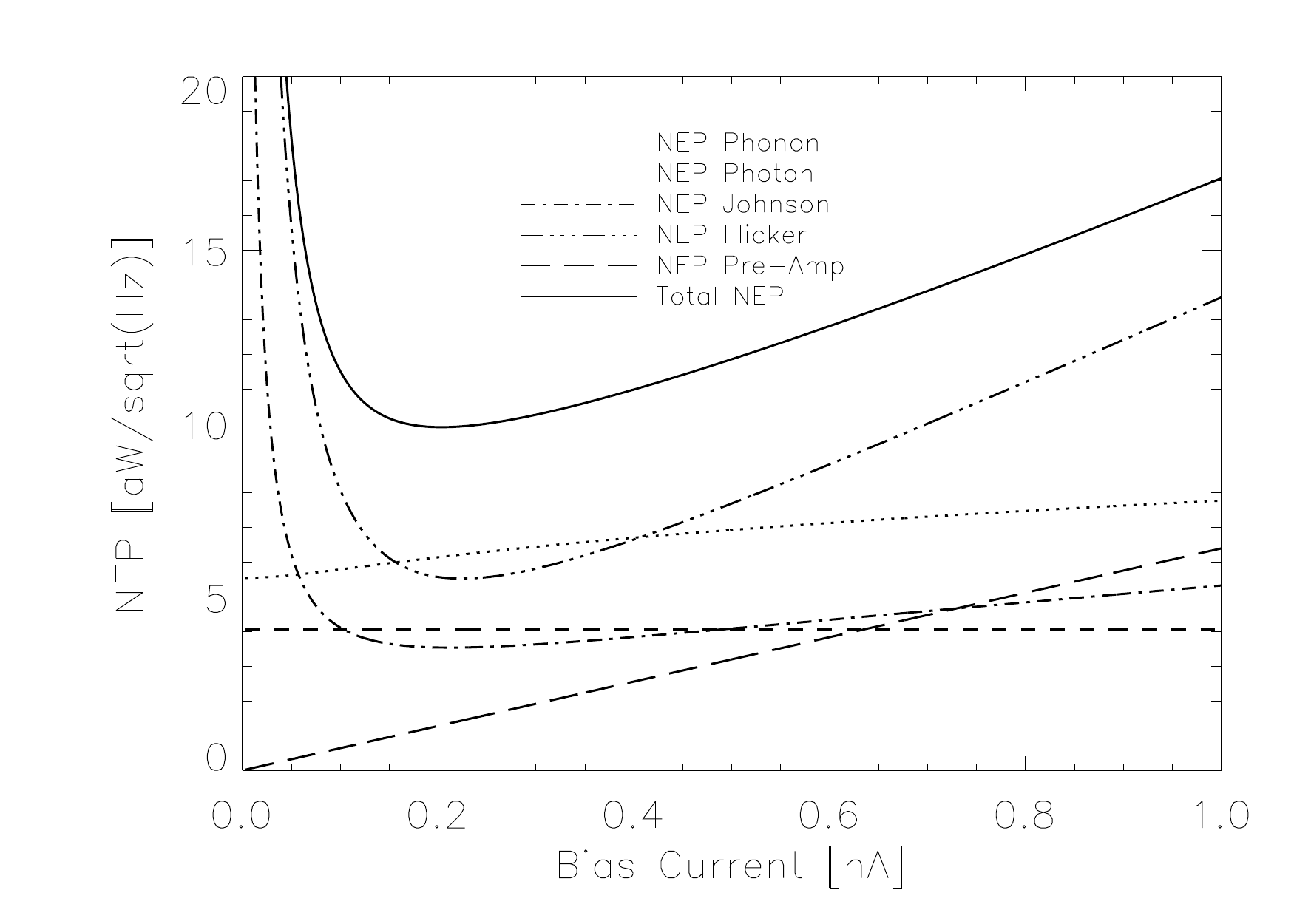}
  \includegraphics[width=8cm,keepaspectratio]{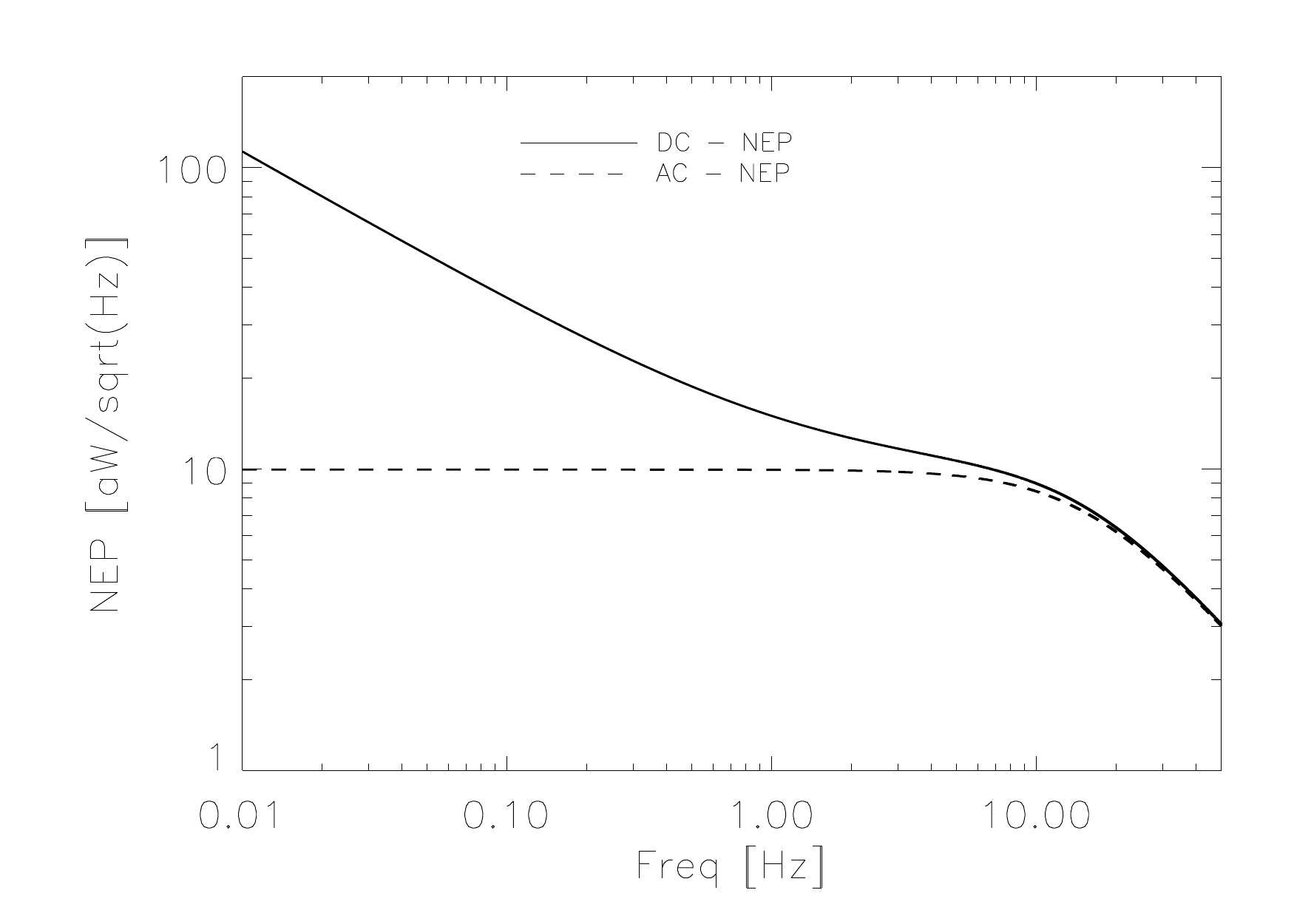}
  \caption{Left: noise equivalent power versus bias current in the bolometer for different sources
    of noise in case of an AC readout electronics for the test bolometer
    $\lambda$ = 3 mm. Right: total
    NEP versus frequency for a DC and AC readout electronics at respective best bias currents.}
  \label{fig:NEPs}
\end{figure}

\section{Noise}

The Noise Equivalent Power (NEP) is:

\begin{equation}\label{nep}
  NEP(f)=\frac{< \Delta S^2(f)>^{1/2}}{\Re(f)}
\end{equation}
where $\Delta S^2$ is the power spectral density of the noise and $\Re$ is the responsivity of the detector.  NEP is
measured in $[W/ Hz^{1/2}]$.

In our model we take into account all the principal sources of noise
in bolometric detection: Johnson noise, phonon noise, photon noise,
Flicker noise and the preamplifier noise.  The following is a review of the NEPs for the
different sources of noise existing in literature (see for instance
\cite{Mather1982,Lamarre1986}) :

\textbf{Johnson noise:} Johnson noise is the electronic noise
generated by the thermal agitation of electrons inside a bolometer at
equilibrium. It has a white noise spectrum.  The NEP for DC biased
bolometers is \cite{Mather1982}:
\begin{equation}\label{johnnoise}
  NEP_{john}=(4k_BT_bR_bI_b^2)^{1/2} \frac{|Z_b+R_b |}{|Z_b-R_b|}
\end{equation}
where $R_b$ is the bolometer resistance and $Z_b$ is its dynamic
impedance.

Let us notice with Mather \cite{Mather1982} that Johnson noise does not
depend on load impedance. Let us assume here that it does not depend
on stray capacitance in the case of AC biased bolometers. Hereafter,
we shall use Eq.~\ref{johnnoise} indifferently with a DC model or an AC
Model.

\textbf{Phonon noise:} The parameters of the bolometer are strongly
dependent on the temperature, so small variations in temperature
inside the bolometer produce a voltage variation at the ends of the
detector.

It results \cite{Mather1982}:
\begin{equation}
  NEP_{phon}=(4k_BGT^2)^{1/2}
\end{equation}

This result is independent of the readout electronics.

\textbf{Photon noise:} The Photon noise comes from the fluctuations of
the incident radiation due to the Bose-Einstein distribution of the
photon emission. The NEP is \cite{Lamarre1986}:
\begin{equation}
  NEP_{phot}=2 \int_{\Delta \nu} h \nu Q_{\nu} d \nu + (1+P^2)
  \int_{\Delta \nu} \Delta(\nu) Q_{\nu}^2 d \nu
\end{equation}
Where $Q_{\nu}$ is the absorbed optical power per unit of frequency,
$\Delta(\nu)$ is the coherence spacial factor (equal to the inverse of the number of
modes; $\Delta (\nu)=1$ if diffraction limited) and $P$ is the polarisation
degree (0 non-polarised 1 polarised).  This noise corresponds to the
limitation in sensitivity of any instrument because it does not
depends on performances of detectors and readout electronics.

\textbf{Flicker noise:} The Flicker noise depends on a
distribution of time constants due to the recombination and generation
phenomena appearing in semiconductors.

This noise shows a spectrum directly proportional to the bias current
and inversly proportional to the frequency. To first order we have :

\begin{equation}
  NEP_{fl} = const  \frac{I_b}{\sqrt{freq}}
\end{equation}

The Flicker noise is usually the dominant source of noise up to few Hertz. In the case of AC electronics, we can choose
the working modulation frequency in order to keep the Flicker noise less then the
photon noise (see Fig.~\ref{fig:NEPs} right).

\textbf{Preamplifier noise:} results from the impossibility to amplify
a signal without adding noise, which is a consequence of the
Heisenberg Uncertainty principle.  It also depends on the available
components and on the design of the amplifier.  We assume that the
power spectrum of signal fluctuation is constant and equal to:
$$< \Delta S^2>_{pre}^{1/2}= const =\sigma_{PA} [V/Hz^{1/2}] $$ The
$NEP_{pre}$ results from Eq.~\ref{nep} as:
\begin{equation}
  NEP_{pre}=\frac{\sigma_{PA}}{\Re}
\end{equation}

The Total NEP of the instrument is:

\begin{equation}
  NEP_{tot} = \left[ NEP_{john}^2 + NEP_{phon}^2 + NEP_{phot}^2 +
    NEP_{fl}^2 + NEP_{pre}^2 \right]^{1/2}
\end{equation}

Fig.~\ref{fig:NEPs} (left) presents the influence of the different
contributions to the total NEP for an AC readout electronics. 

Fig.~\ref{fig:NEPs}  (right) shows the advantage of using of an
AC system instead of a DC solution at low frequencies.
It is clear that Flicker noise increases the DC total NEP at low
frequencies making the AC solution mandatory for the measurement of low and very low frequency signals (less then few Hertz).

\textbf{Optimization of the Bias Current :} from Fig.~\ref{fig:comp}
it is clear that in both cases (AC and DC), the responsivity strongly
depends on the bias current in the bolometer. The
optimisation of this parameter is therefore a key point.  
We want to present a general result, not depending on the preamplifier noise level. Since our practice and all the simulations (see Fig. \ref{fig:NEPs}) show that the minimum NEP happens very near to the maximum responsivity (to better than 1\% in practical cases), we have chosen to use the responsivity to illustrate this point. 
The amplitude and the position of the peak responsivity
are different for the two types of bias current.  In the case
presented in Fig.~\ref{fig:comp}, the bias currents corresponding to
the maximum of the Responsivity are equal to $I^{DC}_{best}$ = 0.12 nA
and $I^{AC}_{best}$ = 0.22 nA.

\section{Variation of Responsivity with Readout Electronics and Environmental Parameters}

We are now interested in establishing the performance of AC readout electronics
biased with a sine wave. We will derive the responsivity, the NEP and how
the NEP depends on the main parameters (stray
capacitance, modulation frequency, optical background and plate
temperature) for three typical bolometers optimized to observe the sky
between 0.3 and 3 mm cooled to a temperature of 100 mK.  In
order to obtain an analytical solution to this problem, we developed a
model in the frequency domain using the Fourier formalism. The results
could be also obtained in the case of a square AC model.  The two
methods are detailed in the appendices A and B.

\begin{figure}[t!]
  \centering
  \includegraphics[width=10cm,keepaspectratio]{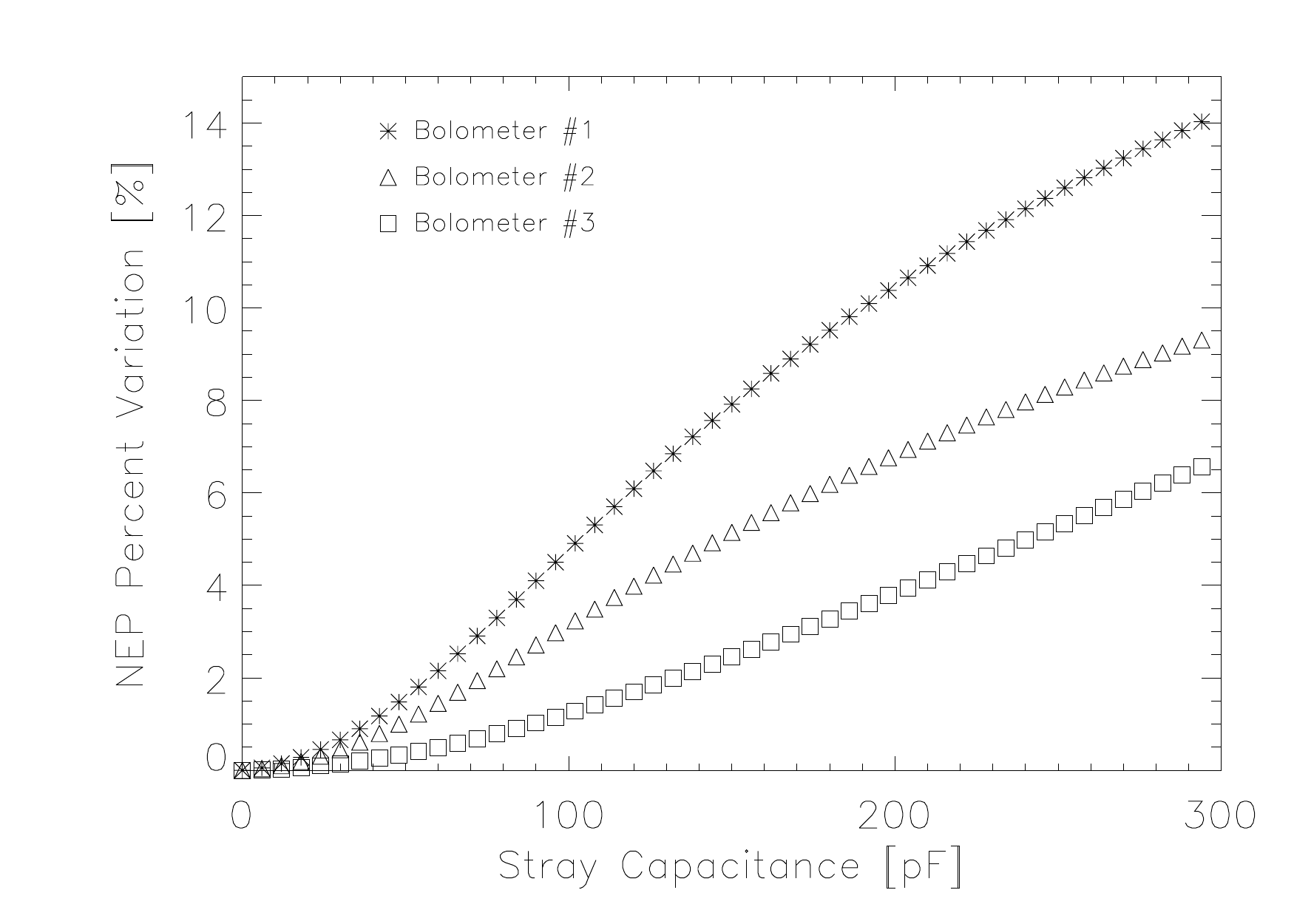}
  \caption{Relative variation of the total NEP versus stray
    capacitance from 0 to 300~pF for three typical bolometers with an
    AC sine bias.}
  \label{fig:nepvscpsin}
\end{figure} 

\begin{figure}[t!]
  \centering
  \includegraphics[width=10cm,keepaspectratio]{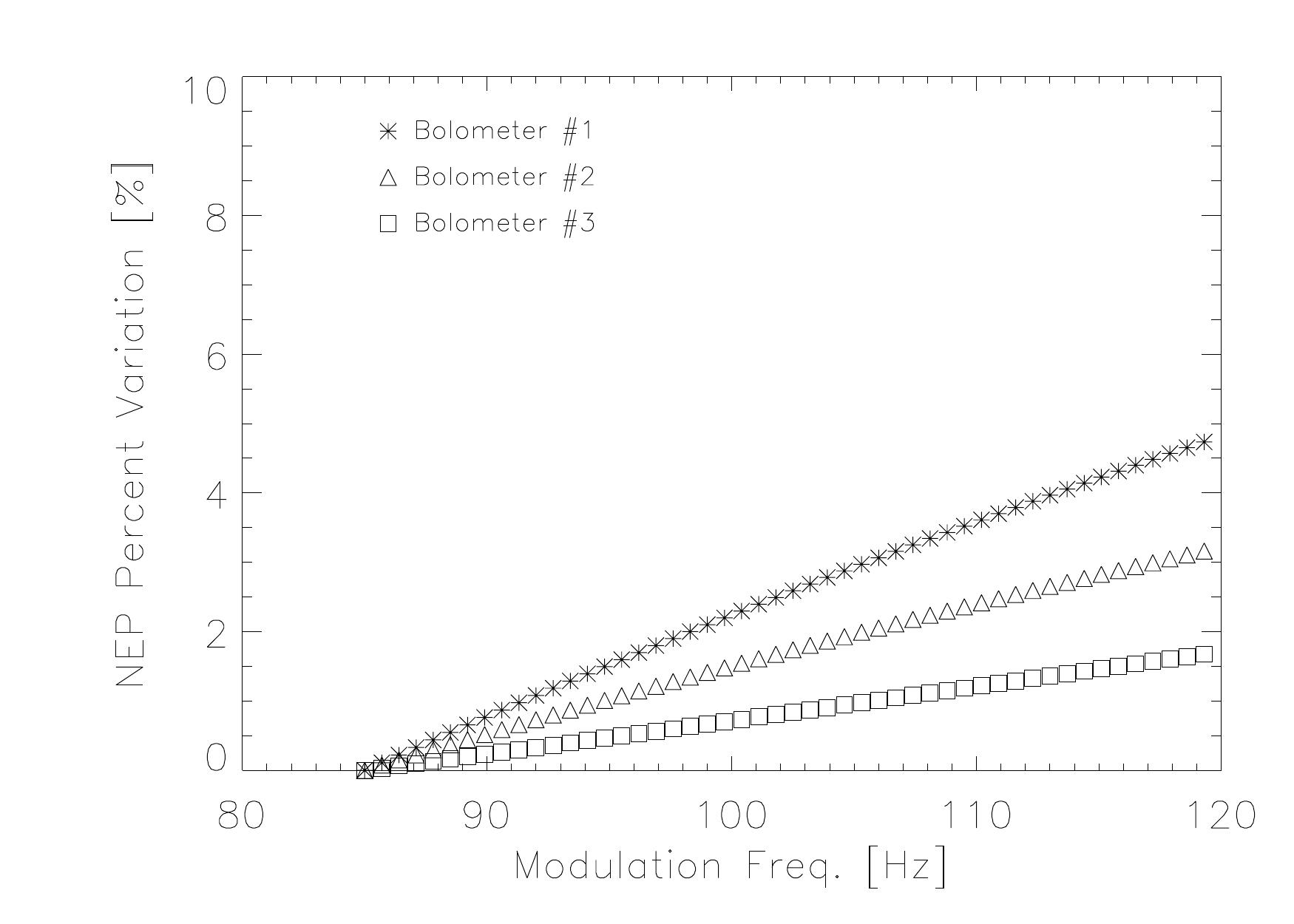}
  \caption{Relative variation of the total NEP versus modulation
    frequency of the AC sine from 85 Hz to 120 Hz for three typical
    bolometers}
\label{fig:nepvsfmodsin}
\end{figure} 

\subsection{Stray Capacitance}

The first stage of preamplifiers for semiconductors bolometers are classically J-FETs giving optimal performance at 100 K or more. Rather long wiring is needed between the J-FETs and the bolometer to avoid an excessive thermal load on the sub-Kelvin stage supporting the bolometer. Stray capacitance of tens and even hundreds of picoFarads result from this design. In Fig. \ref{fig:nepvscpsin}, we plot for our three test bolometers the excess of NEP versus the value of the stray capacitance. For a typical value of 150 pF, the NEP excess is several percents (from 4 \% to 8 \%). Let us note here that the DC bias case is identical to an AC case without stray capacitance.

\subsection{Modulation Frequency}

As we have seen in previous section, the use of an AC bias has the
advantage of presenting a noise spectrum flat down to very low
frequencies, while DC biased readouts show a large 1/f component at
frequencies less than about 10 Hz.  The modulation frequency of the
electronics will be chosen therefore taking into account the
requirement of keeping the Flicker noise less then the Johnson noise
but also taking into account the scanning strategy of the instrument and
the angular responsivity of the optics.  In the case of Planck HFI for
example \cite{Lamarre2010} the full width at half maximum $\delta$ ranges from 5 to 9 arcmin and the scanning speed is 6 degrees per second. So, in the limit of small
angles, the maximum frequency of interest is given by the relation:

\begin{equation}
  f \sim \frac{v_{ang}}{\delta}
\end{equation}

where $f$ is the frequency of the optical modulation.
In Fig.~\ref{fig:nepvsfmodsin}  we consider the excess NEP with respect to a 85 Hz modulation frequency. In the worst case the excess NEP is 0.5 \% per Hertz.

\subsection{Optical Background}

\begin{figure}[t!]
  \centering
   \includegraphics[width=10cm,keepaspectratio]{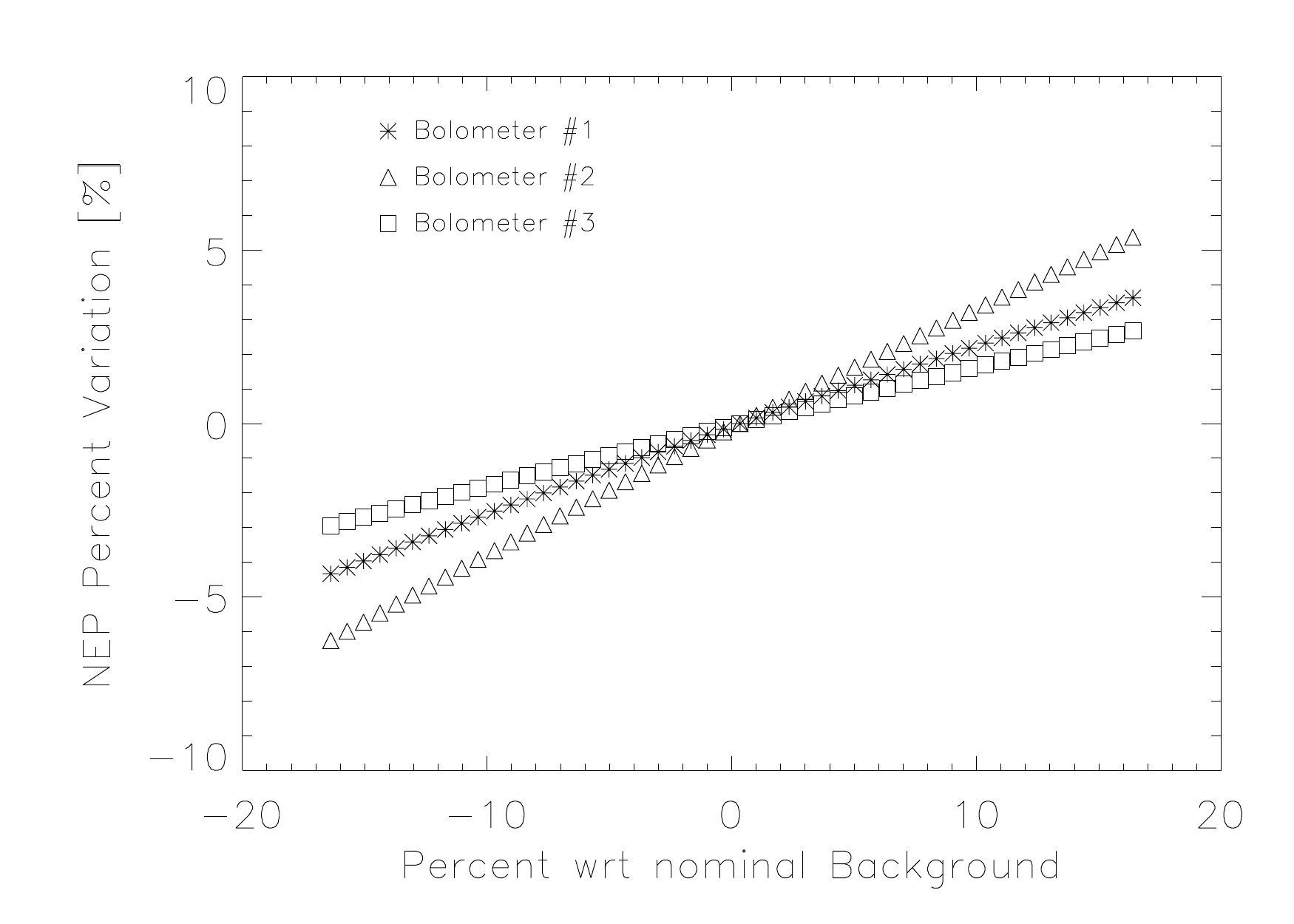}
  \caption{Relative variation of the total NEP versus optical
    background. We consider for each bolometer a total range in background
    equal to 33 \% of the nominal background.}
  \label{fig:nepvswopthfi}
\end{figure}

The background strongly affects the static performance of a bolometer
by changing the operating point. With respect to others
parameters, the background is the most uncertain and variable
parameter during an observational campaign. A good understanding
of the effect of the optical background on the static performances of
a bolometers is therefore a key point during the calibration of the
instrument.  In Fig.~\ref{fig:nepvswopthfi} we present the relative
variation of the total NEP versus the nominal
background for our test bolometers. For the 3 mm
bolometer, the nominal background is 0.3~pW; for 1 mm
0.6~pW and for 0.3 mm 3.6~pW. In the worst case the degradation in
NEP is 2 \% with respect the nominal background for a background increase of +16.5 \%.

\begin{figure}[t!]
  \centering
  \includegraphics[width=10cm,keepaspectratio]{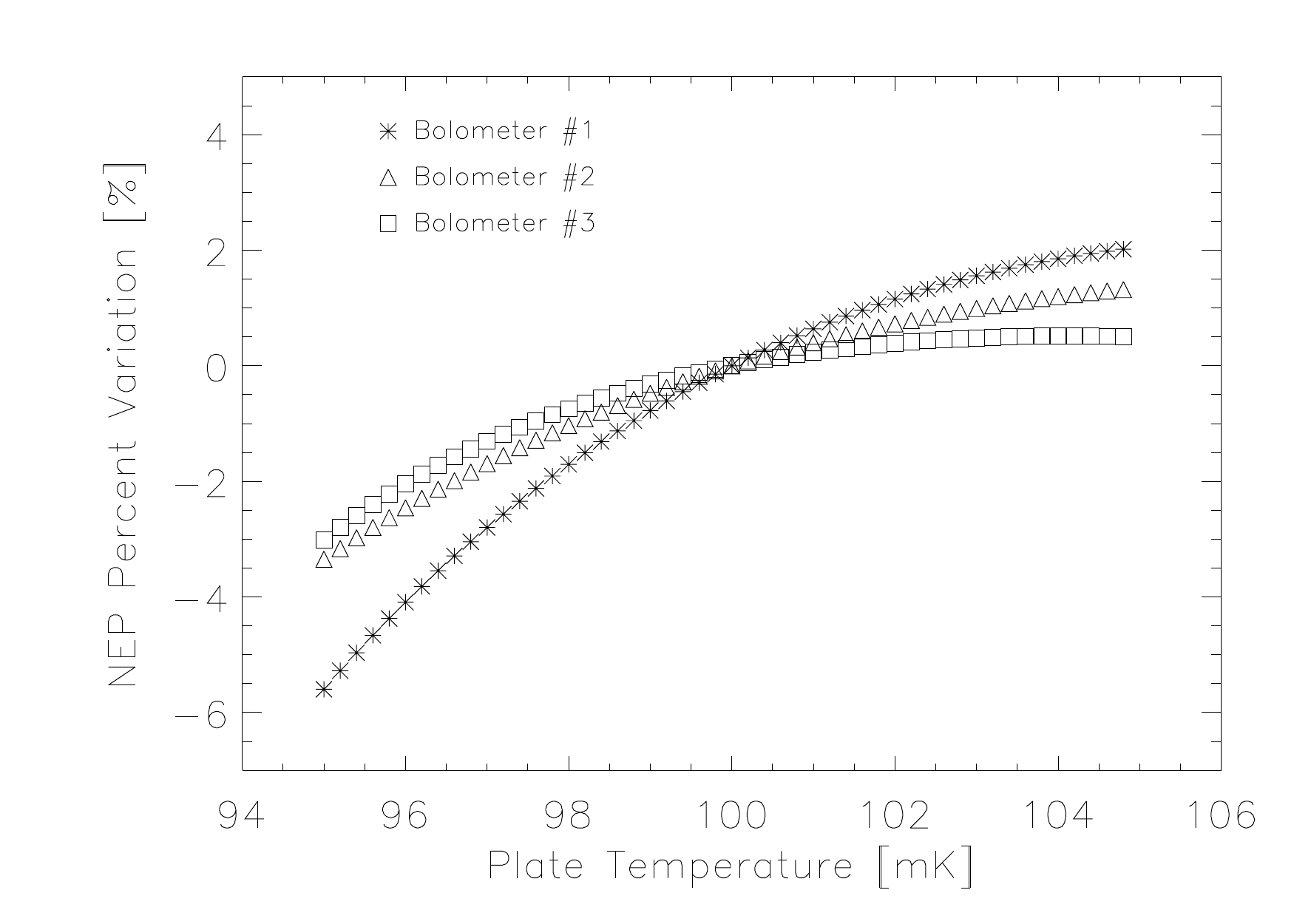}
  \caption{Relative variation of the total NEP versus bolometer plate
    temperature from 95 mK to 105 mK for the three test bolometers.}
  \label{fig:nepvst0hfi}
\end{figure}

\subsection{Bolometer Plate Temperature}

Following the first order thermal model of a bolometer
(Eq.~\ref{bolo}), we know that a change in temperature of the plate
corresponds exactly to a change of background power on the
bolometer. 
In this case the equivalent power generated from a change of the plate
temperature is:

\begin{equation}
  \Delta P_{plate}=G_s \Delta T_0
\end{equation}

Our test bolometers were designed for a plate temperture of 100 mK. The NEP variation in the range 95 mK -- 105 mK is reported in Fig. \ref{fig:nepvst0hfi}.
We find that a change of the plate temperature of 1 mK gives a change
in NEP of 0.8 \% in the worst case.

\section{Comparison Between two Modulation Techniques : Sine AC Bias
  vs Square AC Bias}

\begin{figure}[t!]
  \centering
  \includegraphics[width=8cm,keepaspectratio]{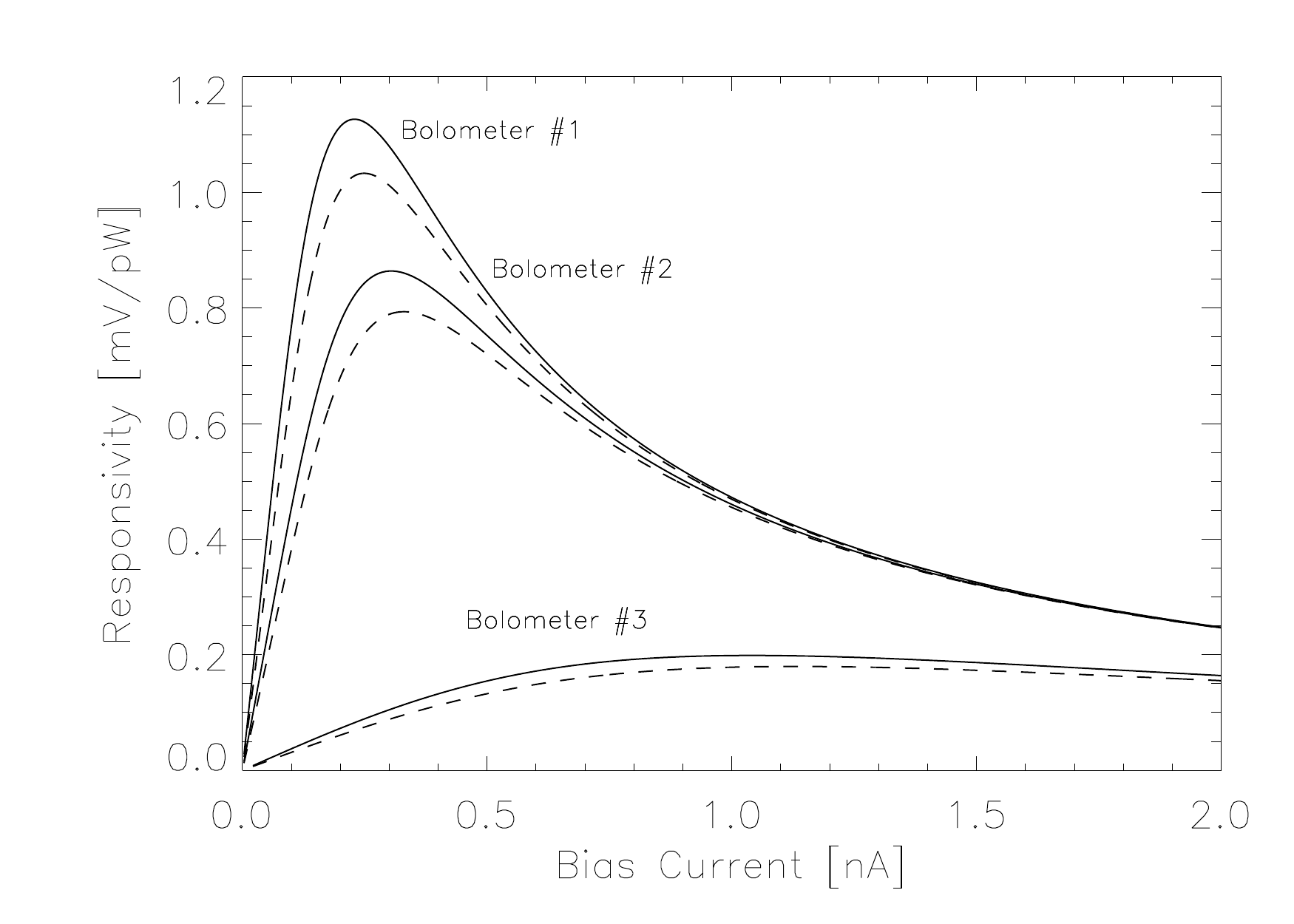}
  \includegraphics[width=8cm,keepaspectratio]{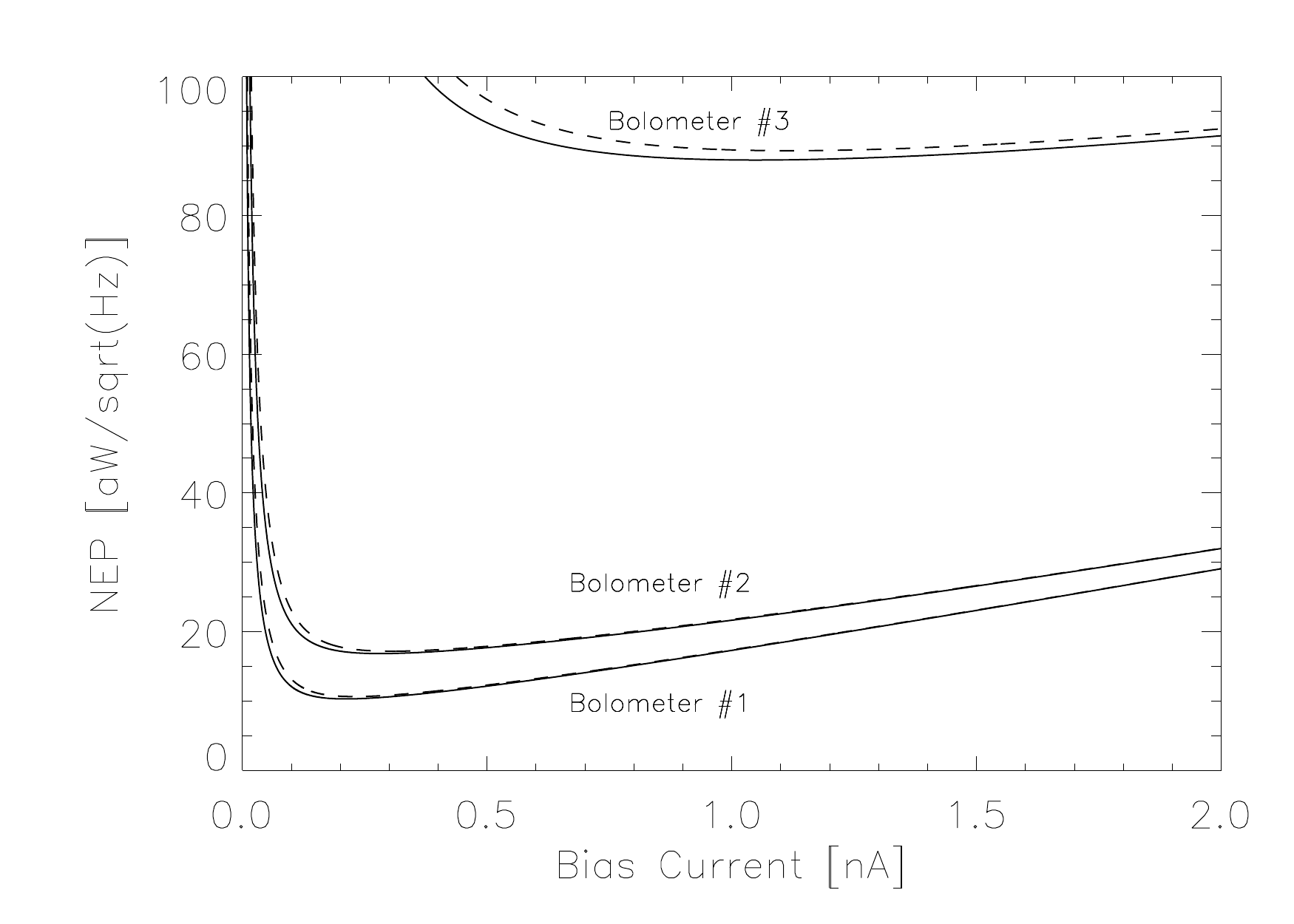}
  \caption{Simulation of responsivity (left) and total NEP (right) of
    the three test bolometers in case of an AC sine bias (solid
    curves) and a square AC bias (dashed curves)}
  \label{fig:nepcomp}
\end{figure}

We want now to compare the performances, in terms of responsivity and NEP,
of a sine-wave and a square-wave AC electronics. The results
are shown in Fig.~\ref{fig:nepcomp}. The sine case is better for
both responsivity and NEP. This is more obvious for the responsivity (better by about 10 \%)
than for the total NEP (better by about 4 \%).  We conclude that
in terms of NEP, a bolometer connected to a sine AC biased readout electronics would be
more sensitive. 

Let us remark that the difference in NEP is modest. Let us also notice
that an AC sine bias would induce significant variations in the
temperature of the fastest bolometers, bringing them into the non-linear
regime.  On the contrary, the square bias deposits a nearly constant
power in the bolometers, that deviate from their mean temperature only
by small amounts.

\section{Conclusion}

  \begin{table}[t!]
    \centering
     \begin{tabular}{lcccccc}
       \hline 
              & wavelength & $R_{*} [Ohm]$  & $G_{so} [pW/K]$ & $T_g [K]$ & $n$&$\beta$  \\
       \hline 
Bolo \#1 & 3 mm   &    100    &     52  &   16   &  0.5 &  1.3\\
Bolo \#2 & 1 mm  &     94    &   70    &  16    &  0.5  &   1.3 \\
Bolo \#3 & 0.3 mm  &    105  &    703   &  16.5    &  0.5  & 1.1\\
\hline 
\end{tabular} 
\caption{\emph{Parameters of the test bolometers used to illustrate the results of the analytical model.}}\label{tab:tab}
  \end{table}

The analytical model presented in this paper has been developed for the HFI on board Planck satellite. It allowed us to predict the responsivity and the noise of semi-conductor bolometers cooled at 100 mK and biased by AC currents in a realistic environment. It sheds some light on the differences between AC and DC biased bolometer and on the different optimal bias currents for these two cases. Three test bolometers rather similar to Planck's ones were used to illustrate our results. Our main conclusions are:

\begin{itemize}
\item
The AC responsivity is always lower than the DC responsivity. This is due to a more effective electro-thermal feedback. The resulting excess of NEP depends on the relative part of the preamplifier noise in the total NEP. In our test cases the excess NEP ranges from 4 \% to 10 \%, which is more than compensated for by shifting of the low frequency noises out of the range of useful frequencies. Frequencies down to 1 mHz are measurable with a well designed AC readout electronics.

\item The AC bias RMS current providing to the maximum of the responsivity is about twice larger than that obtained for a DC bias. This concerns the current through the bolometer and results from the different electro-thermal feedback.

\item For a stray capacitance of $\sim$ 150 pF we obtain an excess NEP of 10 \% in the worst case (3~mm bolometer) and 4\% in the best case (0.3 mm bolometer).

\item Around a modulation frequency of 90 Hz, the excess NEP ranges between 0.2 \% and 0.5 \% per Hz.

\item The sensitivity of NEP to background is dlog(NEP)/dlog(Wbg) = 0.22 to 0.42

\item The sensitivity of NEP to the plate temperature is dlog(NEP)/dlog(Tplate) = 0.3 to 0.8 around 100 mK, but is rather non-linear.

\item The performances of a sine bias are better than the square bias. In our test cases, this result is more obvious in the responsivity (better by about 10 \%) than in the total NEP (better by about 4 \%). But non-linear effects may show up in the sine case for bolometers fast enough to respond to the modulation frequency.
\end{itemize}

\section*{Appendix A:  Computing the Responsivity with a Sine Bias}

Let us consider the bias circuit of Fig.~\ref{fig:ACmod} with a stray
capacitance in parallel to the bolometer and a load capacitance in
series.  The value of the load capacitance is fixed to $C_b=4.7\cdot
10^{-12} F$ which is the typical value in HFI.

Let's also consider a range of temperatures starting from the temperature
of the plate (100mK for example) up to an arbitrary value.  For each
temperature we can calculate the impedance $R_b$ of the bolometer and
its total power using Eqs.  \ref{bolo} and \ref{resi}. In this
simulation we assume that the parameters of the bolometers ($R_*$,
$T_g$, $\beta$, etc......)  are those of HFI.

If the optical background is constant in this run of simulations, the
dissipated electrical power in the bolometer is:
\begin{equation}\label{wele}
  W_{elec}=W_{tot}-W_{opt}
\end{equation}
So, the r.m.s. Voltage at the ends of the bolometer is :
\begin{equation}\label{veff}
  V_b=(R_b W_{elec})^{1/2}
\end{equation}
and the r.m.s. bias current passing through the bolometer is:
\begin{equation}
  I_b=\frac{V_b}{R_b}
\end{equation}

In general for a quadripole we have:
\begin{equation}
  F(V_b)=TF(\omega, R_b, C_p)\cdot F(V_0)
\end{equation}
where $F$ indicate the Fourier transform and $TF(\omega, R_b, C_p)$ is
the transfer function of the quadripole. Using the quadripole obtained
from Eq.~\ref{aceq}, the module of the transfer function is :
\begin{equation}
  |TF(\omega, R_b, C_p)|=\frac{R_b \omega C_b}{(1+\omega^2 R_b^2 (C_b+C_p)^2)^{1/2}}
\end{equation} 
So, the r.m.s. input voltage is:
\begin{equation}
  V_0=\frac{V_b}{|TF(\omega, R_b, C_p)|}
\end{equation}

In order to calculate the optical responsivity let us consider a small
step in temperature for each bolometer. If we
keep $V_0$ unchanged, the step in temperature is due to a change of
the optical background that can be computed:

\begin{equation}
  W_{opt1}=W_{tot1}-W_{elec1}
\end{equation}
where $W_{tot1}$ is calculated from the new temperature $T_{b1}$ and
$W_{elec1}$ is derived from:
\begin{equation}
  W_{elec1}=\frac{V_{b1}^2}{R_{b1}}
\end{equation}
$V_{b1}$ is equal to:
\begin{equation}
  V_{b1}=V_0 \cdot |TF(\omega, R_{b1}, C_p)|
\end{equation}
assuming that $V_0$ is not varying, and using the $TF$ calculated from $R_{b1}$

The responsivity will be:
\begin{equation}
  \Re=| \frac{V_{b1}-V_b}{W_{opt1}-W_{opt}}|
\end{equation}

with the responsivity and NEP equations from the previous section, it is
possible to calculate the total NEP

\section*{Appendix B: Computing the Responsivity with a Square Bias}
With the same bias circuit (Fig.~\ref{fig:ACmod}), it is possible to
derive the performances of a REU in which a square wave voltage applied
to the bolometer, as in HFI. Let us assume that if the REU is
\emph{balanced}, a perfect square wave bias is passing through the
bolometer even in presence of a stray capacitance. In HFI this is
achieved by using a triangular wave plus a square wave.

A square wave can be decomposed as:
$$ V_b(\omega)=a cos (\omega t)+ \frac{a}{3} cos(3 \omega t)
+\frac{a}{5}cos(5 \omega t)+.......=
$$
$$ = \sum_{n=0}^{\bar{n}}\frac{a}{2n+1}cos((2n+1)\omega t)
$$ The r.m.s. $V_{b}$ is equal to:
\begin{equation}\label{hfivb}
  V_{b}=(\sum_{n=0}^{\bar{n}}(\frac{a}{\sqrt{2}(2n+1)})^2)^{1/2}
\end{equation}

If the temperature of the bolometer is given and the optical power is
constant we can calculate the r.m.s. $V_{b}$ as we did for the sine
bias case (Eq.~\ref{veff} and Eq.~\ref{wele}). So we have:
\begin{equation}
  a=(W_{elec} R_b)^{1/2} \cdot \sum_{n=0}^{\bar{n}} 2(2n+1)^2
\end{equation}
The r.m.s. bias current passing through the bolometer is:
\begin{equation}
  I_b=\frac{V_b}{R_b}
\end{equation}

Now let's derive the responsivity. As for the sine AC case, so we can
calculate the $R_{b1}$, $W_{tot1}$, and the $TF$ starting using a small step
in temperature due to an incoming optical signal.

On the other hand we cannot derive the electrical power following the
same logic: if we keep the same set up of the REU, after a small step
in temperature the bias passing through the bolometer is not a square
wave anymore so, the Eq.~\ref{hfivb} is not applicable.  We have to
correct each term of the sum as follows:
\begin{equation}
  W_{elect1}=\frac{a^2}{R_{b1}}\sum_{n=0}^{\bar{n}} 2(2n+1)^2\cdot
  \Upsilon((2n+1)\omega, R_b, R_{b1}, C_p)
\end{equation}
where
\begin{equation}
  \Upsilon((2n+1)\omega, R_b, R_{b1}, C_p)=\frac{TF((2n+1)\cdot\omega,
    R_b, C_p)}{TF((2n+1)\cdot \omega, R_{b1}, C_p)}
\end{equation}


\begin{thebibliography}{10}
\newcommand{\enquote}[1]{``#1''}

\bibitem{Bock2009}
J.~J.~A. {Bock}, Philip, L.~{Armus}, J.~{Bally}, D.~{Benford}, A.~{Cooray},
  M.~{Devlin}, S.~{Dodelson}, D.~{Dowell}, P.~{Goldsmith}, S.~{Golwala},
  S.~{Hanany}, M.~{Harwit}, W.~{Holland}, W.~{Holzapfel}, {Kenyon}, {Matt},
  K.~{Irwin}, E.~{Komatsu}, A. E.~{Lange}, D.~{Leisawitz}, A.~{Lee}, B.~{Mason},
  J.~{Mather}, H.~{Moseley}, S.~{Meyer}, S.~{Myers}, H.~{Nguyen}, V.~{Novosad},
  B.~{Sadoulet}, G.~{Stacey}, S.~{Staggs}, P.~{Richards}, G.~{Wilson},
  M.~{Yun}, and J.~{Zmuidzinas}, \enquote{{Superconducting Detector Arrays for
  Far-Infrared to mm-Wave Astrophysics},} in \enquote{astro2010: The Astronomy
  and Astrophysics Decadal Survey,} , vol. 2010 of \emph{ArXiv Astrophysics
  e-prints} (2009), pp. 45--+.

\bibitem{Jones1953}
R.~C. {Jones}, \enquote{{The general theory of bolometer performance},} Journal
  of the Optical Society of America (1917-1983) \textbf{43}, 1--+ (1953).

\bibitem{Mather1982}
J.~C. {Mather}, \enquote{{Bolometer noise: nonequilibrium theory.}} \ao
  \textbf{21}, 1125--1129 (1982).

\bibitem{Mather1984a}
J.~C. {Mather}, \enquote{{Electrical self-calibraion of nonideal bolometers},}
  \ao \textbf{23}, 3181--3183 (1984).

\bibitem{Mather1984b}
J.~C. {Mather}, \enquote{{Bolometers: ultimate sensitivity, optimization, and
  amplifier coupling},} \ao \textbf{23}, 584--588 (1984).

\bibitem{Rieke1989}
F.~M. {Rieke}, A.~E. {Lange}, J.~W. {Beeman}, and E.~E. {Haller}, \enquote{{An
  AC bridge readout for bolometric detectors},} IEEE Transactions on Nuclear
  Science \textbf{36}, 946--949 (1989).

\bibitem{Wilbanks1990}
T.~{Wilbanks}, M.~{Devlin}, A.~E. {Lange}, J.~W. {Beeman}, and S.~{Sato},
  \enquote{{Improved low frequency stability of bolometric detectors},} IEEE
  Transactions on Nuclear Science \textbf{37}, 566--572 (1990).

\bibitem{Delvin1993}
M.~{Devlin}, A.~E. {Lange}, T.~{Wilbanks}, and S.~{Sato}, \enquote{{A
  dc-coupled, high sensitivity bolometric detector system for the Infrared
  Telescope in Space},} IEEE Transactions on Nuclear Science \textbf{40},
  162--165 (1993).

\bibitem{Gaertner1997}
S.~{Gaertner}, A.~{Beno{\^i}t}, J.-M.~{Lamarre}, M.~{Giard}, J.~{Bret},
  J.~{Chabaud}, F.~{D\'esert}, J.~{Faure}, G.~{Jegoudez}, J.~{Land\'e},
  J.~{Leblanc}, J.~{Lepeltier}, J.~{Narbonne}, M.~{Piat}, R.~{Pons},
  G.~{Serra}, and G.~{Simiand}, \enquote{{A new readout system for bolometers
  with improved low frequency stability},} \aass \textbf{126}, 151--160 (1997).

\bibitem{Kreysa2003}
E.~{Kreysa}, F.~{Bertoldi}, H.~{Gemuend}, K.~M. {Menten}, D.~{Muders}, L.~A.
  {Reichertz}, P.~{Schilke}, R.~{Chini}, R.~{Lemke}, T.~{May}, H.~{Meyer}, and
  V.~{Zakosarenko}, \enquote{{LABOCA: a first generation bolometer camera for
  APEX},} in \enquote{Society of Photo-Optical Instrumentation Engineers (SPIE)
  Conference Series,} , vol. 4855 of \emph{Society of Photo-Optical
  Instrumentation Engineers (SPIE) Conference Series}, {T.~G.~Phillips \&
  J.~Zmuidzinas}, ed. (2003), vol. 4855 of \emph{Society of Photo-Optical
  Instrumentation Engineers (SPIE) Conference Series}, pp. 41--48.

\bibitem{Lamarre2010}
J.-M. {Lamarre} et al., \enquote{{Planck pre-launch status: the HFI instrument, from specifications to actual performance}}, in press, \aa  \ (2010).

\bibitem{Vaillancourt2005}
J.~E. {Vaillancourt}, \enquote{{Complex impedance as a diagnostic tool for
  characterizing thermal detectors},} Review of Scientific Instruments
  \textbf{76}, 043107--+ (2005).

\bibitem{Piat2006}
M.~{Piat}, J.-P.~{Torre}, E.~{Br{\'e}elle}, A.~{Coulais}, A.~{Woodcraft},
  W.~{Holmes}, and R.~{Sudiwala}, \enquote{{Modeling of Planck-high frequency
  instrument bolometers using non-linear effects in the thermometers},} Nuclear
  Instruments and Methods in Physics Research A \textbf{559}, 588--590 (2006).

\bibitem{Zwerdling1968}
S.~{Zwerdling}, \enquote{{A fast, high-responsivity bolometer detector for the
  very-far infrared},} Infrared Physics \textbf{8}, 271--336 (1968).

\bibitem{Lamarre1986}
J.-M. {Lamarre}, \enquote{{Photon noise in photometric instruments at
  far-infrared and submillimeter wavelengths},} \ao \textbf{25}, 870--876
  (1986).

\end{thebibliography}
\end{document}